\documentclass{article}
\usepackage[margin=1in]{geometry}
\usepackage{amsmath}
\usepackage{epsfig}
\numberwithin{equation}{section} % amsmath

\usepackage{tcolorbox}
\usepackage{pict2e}
\newcommand{\beq}{\begin{equation}}
\newcommand{\eeq}{\end{equation}}
\newcommand{\beqa}{\begin{eqnarray}}
\newcommand{\eeqa}{\end{eqnarray}}
\newcommand{\bdm}{\begin{displaymath}}
\newcommand{\edm}{\end{displaymath}}
\newcommand{\lslash}[1]{#1\llap/}

\newcommand{\Eq}[1]{Eq.\ (\ref{#1})}
\newcommand{\Eqs}[2]{Eqs.\ (\ref{#1}) and (\ref{#2})}

\newcommand{\Rref}[1]{Ref.\ \cite{#1}}
\newcommand{\Rrefs}[2]{Refs.\ \cite{#1} and \cite{#2}}

\newcommand{\Fig}[1]{Fig.\ \ref{#1}}

\newcommand{\Section}[1]{Section\ \ref{#1}}
\newcommand{\Appendix}[1]{Appendix\ \ref{#1}}
\title{Model for the propagation of fermions in a Bose-Einstein condensate}

\author{Jos\'e F. Nieves\footnote{nieves@ltp.uprrp.edu}\\
  Laboratory of Theoretical Physics, Department of Physics\\
  University of Puerto Rico, R\'{\i}o Piedras, Puerto Rico 00936
  \and\\[12pt]
  Sarira Sahu\footnote{sarira@nucleares.unam.mx}\\
  Instituto de Ciencias Nucleares\\
  Universidad Nacional Aut\'onoma de Mexico\\
  Circuito Exterior, C. U.\\
  A. Postal 70-543, 04510 Mexico DF, Mexico\\
}

\date{November 2022}

\begin{document}
\maketitle

\begin{abstract}
  We consider the propagation of fermions in the
  background of a scalar Bose-Einstein condensate.
  Some illustrative examples are discussed using simple Yukawa-type
  coupling models between the fermions and the scalar fields.
  The fermion dispersion relations are determined explicitly
  in those cases, to the lowest order, and in each case we
  discuss some of the properties of the propagating fermion modes.
  We also obtain the dispersion relations and wavefunctions of the scalar
  modes, which can be used to obtain the corrections (e.g., damping effects)
  to the fermion dispersion relations due to the interactions with
  the excitations of the Bose-Einstein condensate. Possible applications
  of these results in some contexts, such as neutrinos propagating in a scalar
  Dark Matter background, are mentioned.
\end{abstract}

\section{Introduction and motivation}
\label{sec:intro}

In several models and extensions of the standard electroweak
theory the neutrinos interact with a scalar ($\phi$)
and fermion ($f$) via a coupling of the form $\bar f_R\nu_L\phi$, or just
with neutrinos themselves $\bar\nu^c_R\nu_L\phi$. 
Couplings of the form $\bar f_R\nu_L\phi$ produce additional contributions to
the neutrino effective potential when the neutrino propagates
in a background of $\phi$ and $f$ particles
and their possible effects have been considered in
various contexts, such as collective oscillations in supernova
(see for example \Rrefs{Duan:2010bg}{Chakraborty:2016yeg}
and the works cited therein), the hot plasma of the
Early-Universe\cite{Wong:2002fa,Mangano:2006ar},
cosmological observations such as cosmic microwave background
and big bang nucleosynthesis data\cite{babu}, and
in particular  Dark Matter-neutrino interactions\cite{Mangano:2006mp,
Binder:2016pnr,Primulando:2017kxf,Campo:2017nwh,Franarin:2018gfk,
Pandey:2018wvh}.

Motivated by these developments, we have carried out in
previous works a systematic calculation of the neutrino
dispersion relation in such models, including the damping
and decoherence effects (see \Rref{ns:nuphiresonance} and references therein).
These works have been based on the calculation of the neutrino
thermal self-energy using thermal field theory (TFT) methods\cite{ftft:reviews}.

Analytic formulas for the various quantities of interest have been
obtained by considering various different cases of the
$f$ and $\phi$ background, such as the non-relativistic
or ultra-relativistic gases, and in particular the case in which the
$f$ background is a completely degenerate Fermi-gas.

To complement that previous work, our goal is to determine
the corresponding quantities (e.g, effective potential and/or
dispersion relation and damping) of a neutrino that propagates
in a thermal background that contains a scalar Bose-Einstein (BE) condensate.
The hypothesis that the dark matter (DM)
can be self-interacting is intriguing, and a DM background of
scalar particles is a candidate for such environments\cite{garani:becdm,
kirkpatrick:becdm,bohmen:becdm,craciun:bdcdm}.
In that context, the interest is the application
to the case of a neutrino propagating in such a background.

The problem of fermions propagating in such backgrounds can be relevant
in other contexts as well. For example, the possibility of BE condensation of
pions and/or kaons in the interior of a neutron star,
or kaon condensation in heavy ion collisions\cite{baym:nstar,
thorsson:kaon,schmitt:kaon,li:hion}.

Our purpose here is to propose an efficient and consistent method to treat
the propagation of a fermion in the background of the BE condensate,
in particular the calculation of the effective potential
and dispersion relation, in a general way and not tied to any
specific application. To model the fermion propagation in
such an environment, we assume some simple Yukawa-type interactions
between the fermions and the scalar.

We consider three generic, but specific, models of the fermion-scalar
interaction:
\begin{enumerate}
\item Model I: Two massless chiral fermions, $f_L$ and $f_R$,
  with a coupling to the scalar particle $\phi$ of the form
  $\bar f_R f_L\phi$.
\item Model II: A massless chiral fermion $f_L$ with coupling
  $\bar f^c_R f_L\phi$.
\item Model III: One massive Dirac fermion $f$ with a coupling
  $\bar f^c f\phi$.
\end{enumerate}
As we will see, the symmetry breaking process produces
a Dirac fermion, a Majorana fermion and a pseudo-Dirac fermion
in Model I, II and III, respectively\cite{wolfensteinpetcov}.

The field theoretical method we use to treat the BE condensate
has been discussed by various authors\cite{weldon:phimu,filippi:phimu,
schmitt:phimu}. For completeness we first discuss those aspects and
details of the method that are relevant for our purposes.
We then present the extension we propose to treat the fermion propagation in
the BE condensate, in the context of the three models mentioned above
for concreteness and illustrative purposes. Although one of our motivations
is the possible application in neutrino physics contexts,
the method we propose for the propagation of fermions in a BE condensate
has never being used before, and most importantly, is general and paves the way
for applications to problems in other systems, for example
condensed matter, or nuclear matter systems and
heavy-ion collisions as already mentioned.

The plan of the paper is as follows. In \Section{sec:becmodel} we review
the model we use to describe the BE condensate.
There we focus on the essential elements of the symmetry breaking
mechanism that we need in the next sections.
In \Section{sec:modeli} we consider in detail the method we use for
calculating the dispersion relations of the propagating fermions in
the BE condensate, in the context of the model-I mentioned above.
The method is further illustrated by applying it to the models II and III
in Sections \ref{sec:modelii} and \ref{sec:modeliii}, respectively.

With a view to possible interest and/or future work, we summarize
in an appendix the details related to the scalar modes that have a
definite dispersion relation, which are useful for the calculation
of the thermal corrections to the fermion dispersion relations due
to the thermal excitations of the BE condensate.
Our concluding remarks and outlook are given in \Section{sec:conclusions}.

\section{Model for the BE condensate}
\label{sec:becmodel}

To describe the BE condensate the proposal is to start with
the complex scalar field $\phi$ that has a standard $\phi^4$ Lagrangian
\beq
\label{Lphi}
L^{(\phi)} =  (\partial^\mu\phi)^\ast(\partial_\mu\phi) - V_0\,,
\eeq
where
\beq
\label{V0}
V_0(\phi) = m^2_\phi\phi^\ast\phi + \lambda_\phi(\phi^\ast\phi)^2\,.
\eeq
Critical examinations in the literature (see, e.g., \Rref{guth})
support the notion that this potential can indeed lead to thermalization
and formation of a stable condensate due to repulsive interactions, that can
drive long-range order, for $\lambda_\phi > 0$
(as opposed to $\lambda_\phi < 0$). Thus, for definiteness,
here we assume that
\beq
\lambda_\phi > 0\,,
\eeq
so that this condition to allow forming a stable condensate is satisfied.

In the context of thermal field theory (TFT), denoting the temperature by $T$
and the chemical potential of $\phi$ by $\mu_\phi$,
the procedure is to calculate the \emph{effective potential} of $\phi$, call it
$V^{(\phi)}_\text{eff}(T,\mu_\phi)$, and then see under what conditions
$V^{(\phi)}_\text{eff}$ has minimum at $\phi = 0$
or some other value. In the latter case, there has been a phase transition,
and
\beq
\langle\phi\rangle \not = 0\,,
\eeq
indicative of the symmetry breaking.

The alternative approach that we use, which is particularly useful for
treating the symmetry breaking associated with the transition to the
BE condensate, is to consider the field $\phi^\prime$ defined
by\cite{weldon:phimu,filippi:phimu,schmitt:phimu}
\beq
\label{defphiprime}
\phi^\prime \equiv e^{i\mu_\phi t}\phi\,.
\eeq
The recipe is to substitute $\phi = e^{-i\mu_\phi t}\phi^\prime$
in $L^{(\phi)}$ to obtain the Lagrangian for the field $\phi^\prime$,
which we denote by $L^{(\phi^\prime)}$.
To express $L^{(\phi^\prime)}$ in a convenient form we write
\beq
\mu_\phi t = \mu_\phi (u\cdot x)\,,
\eeq
where
\beq
\label{u}
u^\mu = (1,\vec 0)\,,
\eeq
and define
\beq
\label{D}
D_\mu \equiv \partial_\mu - i v_\mu\,,
\eeq
with
\beq
\label{v}
v_\mu = \mu_\phi u_\mu\,.
\eeq
Then using
\beq
\label{derivativerelation}
\partial_\mu\phi =
\partial_\mu(e^{-i\mu_\phi t}\phi^\prime) = e^{-i\mu_\phi t}D_\mu\phi^\prime\,,
\eeq
it follows that
\beq
\label{Lphiprime}
L^{(\phi^\prime)} = 
(D^\mu\phi^\prime)^\ast(D_\mu\phi^\prime) - V_0(\phi^\prime)\,.
\eeq
Expanding the $D$ term in \Eq{Lphiprime},
\beq
\label{Lmu}
L^{(\phi^\prime)} = (\partial^\mu\phi^\prime)^\ast(\partial_\mu\phi^\prime) +
i[\phi^{\prime\,\ast} (v\cdot\partial\phi^\prime) -
(v\cdot\partial\phi^\prime)^\ast \phi^\prime] - U(\phi^\prime)\,,
\eeq
where
\beq
\label{U}
U = -(\mu^2_\phi - m^2_\phi)\phi^{\prime\,\ast}\phi^\prime +
\lambda_\phi(\phi^{\prime\,\ast}\phi^\prime)^2\,.
\eeq
Now comes the key observation. If $m^2_\phi > \mu^2_\phi$, this $U$ corresponds
to a standard massive complex scalar with mass $m^2_\phi - \mu^2_\phi$.
On the other hand, if $\mu^2_\phi > m^2_\phi$, the minimum of the potential is
not at $\phi = 0$, and therefore $\phi$ develops a non-zero expectation
value and the $U(1)$ symmetry is broken.

We assume the second option,
\beq
\label{sbcondition}
\mu^2_\phi > m^2_\phi\,,
\eeq
and proceed accordingly. Namely, we put
\beq
\label{phi}
\phi^\prime = \frac{1}{\sqrt{2}}\left(\phi_0 + \phi_1 + i\phi_2\right),
\eeq
where
\beq
\label{phiexpectationvalue}
\langle\phi^\prime\rangle \equiv \frac{1}{\sqrt{2}}\phi_0\,,
\eeq
is chosen to be the minimum of
\beq
\label{U0}
U_0 = -\frac{1}{2}(\mu^2_\phi - m^2_\phi)\phi^2_0 +
\frac{1}{4}\lambda_\phi\phi^4_0\,.
\eeq
Thus,
\beq
\label{phi0}
\phi^2_0 = \frac{\mu^2_\phi - m^2_\phi}{\lambda_\phi}\,.
\eeq
Substituting \Eqs{phi}{phi0} in \Eq{Lphi} we obtain the
Lagrangian for $\phi_{1,2}$. $\phi_1$ and $\phi_2$ are mixed
by the $v^\mu$ term.

The central result that we invoke now is that the calculation of the effective
potential $V^{(\phi)}_\text{eff}(T,\mu_\phi)$ can be carried out in TFT using
$\mu_\phi = 0$ in the partition (and/or distribution) function, but using the
$\mu_\phi$-dependent Lagrangian $L^{(\phi^\prime)}$ given in \Eq{Lmu}.
An exhaustive exposition of the equivalence of
using this scheme for the calculation of the effective potential,
or in fact any other physical quantity involving the scalar field,
is given by Weldon\cite{weldon:phimu}. In \Appendix{sec:chemicalpotential}
we give a simplified but precise statement of the arguments involved, and which
further shows the validity to proceed in the same way with the fermion
fields as well. Therefore, following this scheme, the next step would be to
find the propagator matrix of the $\phi_{1,2}$ system,
determine the modes that have a definite dispersion relation,
and then define the thermal propagators of the modes.

However, for our purposes in what follows, it is sufficient to
observe that, neglecting the $T$-dependent terms (that is, at zero temperature),
$V^{(\phi)}_\text{eff}(0,\mu_\phi)$ is simply the $U$ potential given in \Eq{U},
and the zero-temperature expectation value of $\phi^\prime$ is given
by \Eqs{phiexpectationvalue}{phi0}. As we will see, this strategy will allow
us to determine the contribution to the effective potential of fermions
propagating in the BE condensate. The thermal propagators
of the $\phi_{1,2}$ modes would allow us to calculate the
corresponding corrections due to the thermal excitations.
While we do not purse here the calculation of those
thermal corrections, for completeness and possible relevance in future work
we give in \Appendix{sec:phi12modes} some details about the propagator matrix
of the $\phi_{1,2}$ complex, the modes that have a definite dispersion relation,
and the corresponding propagators of the modes.

\section{Model I}
\label{sec:modeli}

\subsection{Formulation}

We consider two chiral fermions $f_L$ and $f_R$, with
an interaction
\beq
L_{\text{int}} = -\lambda \phi \bar f_R f_L + h.c\,.
\eeq
There are two conserved charges, which we will label as $Q_{1,2}$.
The assignments must satisfy
\beq
\label{chargeassignments}
Q_A(\phi) + Q_A(f_L) - Q_A(f_R) = 0\,.
\eeq
We can take
\beqa
Q_1(f_L) = Q_1(f_R) = 1,& \qquad & Q_1(\phi) = 0\,,\nonumber\\
Q_2(\phi) = Q_2(f_R) = 1, & \qquad & Q_2(f_L) = 0\,.
\eeqa
Remembering how the $Q_A$ enter in the partition function operator, namely
\beq
\label{Zgen}
Z = e^{-\beta H + \sum_A\alpha_A Q_A}\,,
\eeq
the assignments in \Eq{chargeassignments} imply that the chemical
potentials satisfy
\beq
\label{murelation}
\mu_\phi + \mu_L - \mu_R = 0\,,
\eeq
where we are denoting by $\mu_L$ and $\mu_R$ the
chemical potential of $f_L$ and $f_R$, respectively.
From our discussion of the BE condensate model in \Section{sec:becmodel}
we take that we should rewrite the Lagrangian in terms of the field
$\phi^\prime$ defined in \Eq{defphiprime}. The generalization
that we propose here is that every field with non-zero $Q_A$
must be transformed accordingly. Therefore, a generalization of the
transformation considered in \Section{sec:becmodel} is to put
\beqa
\label{primedfieldsi}
\phi & = & e^{-i\mu_\phi t}\phi^\prime\,,\nonumber\\
f_L & = & e^{-i\mu_L t}f^\prime_L\,,\nonumber\\
f_R & = & e^{-i\mu_R t}f^\prime_R\,.
\eeqa
From the discussion in \Appendix{sec:chemicalpotential}, it follows
that we can use the partition function given by \Eq{Znomu},
without the chemical potential, uniformly for all the
fields involved, provided we also use the dynamical
equations that follow from the transformed Hamiltonian or Lagrangian.
In short, our proposal here is that the prime fields,
$f^\prime_R$ and $f^\prime_L$, are the appropriate
ones to use to determine the fermion modes in
the BE condensate.

With the condition in \Eq{murelation}, the interaction coupling
keeps the same form, namely
\beq
\label{Lintprime}
L_{\text{int}} = -\lambda \phi^\prime \bar f^\prime_R f^\prime_L + h.c\,.
\eeq
However, the kinetic part of the Lagrangian changes. For $\phi^\prime$
we will borrow what we did in \Section{sec:becmodel}. But now we have to do
something analogous for the fermion fields.

The kinetic part of the fermion Lagrangian,
\beq
L_{f} = i\bar f_L\lslash{\partial} f_L +
i\bar f_R\lslash{\partial}f_R\,,
\eeq
in terms of $f^\prime_R$ and $f^\prime_L$ is
\beq
L_{f} = i\bar f^\prime_L\lslash{\partial}f^\prime_L +
i\bar f^\prime_R\lslash{\partial}f^\prime_R +
\mu_L\bar f^\prime_L\lslash{u}f^\prime_L +
\mu_R\bar f^\prime_R\lslash{u}f^\prime_R\,.
\eeq
As discussed in \Section{sec:becmodel}, we assume a symmetry
breaking by the mechanism implemented around \Eq{sbcondition}.
Therefore, we put
\beq
\langle\phi^\prime\rangle \equiv \frac{1}{\sqrt{2}}\phi_0\,,
\eeq
where $\phi_0$ is given in \Eq{phi0}.
As a result $Q_2$ is broken, but $Q_1$ remains unbroken.
This produces a mass term in \Eq{Lintprime} of the form
\beq
-m\bar f^\prime_R f^\prime_L + h.c.\,,
\eeq
with
\beqa
\label{msymmbreaking}
m & = & \frac{\lambda\phi_0}{\sqrt{2}}\,,\nonumber\\
& = & \frac{\lambda}{\sqrt{2}}\left(
\frac{\mu^2_\phi - m^2_\phi}{\lambda_\phi}\right)^{1/2}\,,
\eeqa
where in the second equality we have used \Eq{phi0}.
The total Lagrangian is then
\beq
\label{modelitotalL}
L = L^{(\phi^\prime)} + L_0 + L^\prime_{\text{int}}\,,
\eeq
where $L^{(\phi^\prime)}$ is given in \Eq{Lmu},
\beq
L_0 = \bar f^\prime_L i\lslash{\partial} f^\prime_L + 
\bar f^\prime_R i\lslash{\partial} f^\prime_R +
\mu_L\bar f^\prime_L\lslash{u} f^\prime_L +
\mu_R \bar f^\prime_R \lslash{u} f^\prime_R -
(m\bar f^\prime_R f^\prime_L + h.c.)\,,
\eeq
and
\beq
L^\prime_{\text{int}} = -\frac{\lambda}{\sqrt{2}}
  (\phi_1 + i\phi_2) \bar f^\prime_R f^\prime_L + h.c\,.
\eeq

Defining
\beq
f = f^\prime_L + f^\prime_R\,,
\eeq
in momentum space $L_0$ is given by
\beq
L_0(k) = \bar f(k)(\lslash{k} - \Sigma(k))f(k)\,,
\eeq
where
\beq
\label{Sigmai}
\Sigma = mL + m^\ast R - \mu_L\lslash{u}L - \mu_R\lslash{u}R \,.
\eeq
The two chiral fermions form a Dirac particle, in which the left and right
components have different dispersion relations. The next step
is to find the propagating modes (dispersion relations and wave functions)
at the tree-level. This is most conveniently done using the
Weyl representation of the $\gamma$ matrices.

\subsection{Dispersion relations}

The field equation in momentum space is
\beq
(\lslash{k} - \Sigma)f = 0\,,
\eeq
or, in terms of the left- and right-hand components of $f$,
\beqa
\lslash{A}_L f^\prime_L - m^\ast f^\prime_R & = & 0\nonumber\\
\lslash{A}_R f^\prime_R - m f^\prime_L & = & 0\,,
\eeqa
where
\beqa
A_{L\mu} & = & k_\mu + \mu_L u_\mu\nonumber\\
A_{R\mu} & = & k_\mu + \mu_R u_\mu\,.
\eeqa
In the one-generation case we are considering the
phase of $m$ is irrelevant, since it can be absorbed by a field redefinition,
so that we could take $m^\ast = m$. However, since in more general
cases such field redefinitions cannot be done independently,
we keep $m$ arbitrary.

We use the Weyl representation of the gamma matrices and put
\beqa
f^\prime_L & = & \left(\begin{array}{cc}0\\ \eta\end{array}\right)\,,
  \nonumber\\
f^\prime_R & = & \left(\begin{array}{cc}\xi\\ 0\end{array}\right)\,.
\eeqa
The equations to be solved then become
\beqa
\label{modelixietaeqs}
\left(A_L^0 + \vec{\sigma}\cdot\vec\kappa\right)\eta - m^\ast\xi & = & 0\,,
\nonumber\\
\left(A_R^0 - \vec{\sigma}\cdot\vec\kappa\right)\xi - m\eta & = & 0\,,
\eeqa
where
\beqa
A_{L0} & = & \omega + \mu_L\,,\nonumber\\
A_{R0} & = & \omega + \mu_R\,,
\eeqa
and we have used $\vec A_{L} = \vec A_R = \vec\kappa$. In general,
leaving out the case that $\mu_R = \mu_L$
(i.e., assuming $\mu_\phi \not= 0$),
these equations have non-trivial solutions only if $\xi$ and $\eta$ are
proportional to the same eigenvector of $\vec{\sigma}\cdot\vec\kappa$.
This can be seen in various ways. For example, using the second equation
of \Eq{modelixietaeqs} to eliminate $\eta$ in the first equation gives
\beq
\left[A^0_L A^0_R - \kappa^2  - |m|^2 +
  \vec\sigma\cdot\vec\kappa(A^0_R - A^0_L)\right]\xi = 0\,,
\eeq
which implies that $\xi$ is eigenvector of $\vec{\sigma}\cdot\vec\kappa$,
and then the second equation implies that $\eta$ is proportional to $\xi$.

Therefore, we write the solution in the form
\beqa
\label{etachiparammodeli}
\eta & = & x\chi_s\,,\nonumber\\
\xi & = & y\chi_s\,,
\eeqa
where $\chi_s$ is the spinor with definite helicity, defined by
\beq
\label{helicityspinor}
\left(\vec\sigma\cdot\hat\kappa\right)\chi_s = s\chi_s\,,
\eeq
with $s = \pm 1$. For a given helicity $s$, the equations for $x$ and $y$ are
\beqa
\label{xymodeli}
(\omega + s\kappa + \mu_L)x - m^\ast y & = & 0\,,\nonumber\\
(\omega - s\kappa + \mu_R)y - mx & = & 0\,,
\eeqa
which imply that $\omega$ must satisfy
\beq
(\omega + s\kappa + \mu_L)(\omega - s\kappa + \mu_R) - |m|^2 = 0\,.
\eeq
Expressing $\mu_R$ and $\mu_L$ in terms of their sum and their difference
$\mu_R \pm \mu_L$, this equation can be written in the form
\beq
\left[\omega + \frac{1}{2}(\mu_R + \mu_L)\right]^2 -
\left[s\kappa - \frac{1}{2}(\mu_R - \mu_L)\right]^2 - |m|^2 = 0\,.
\eeq

For each $s$, we have two solutions, one with positive $\omega$ and
another with a negative $\omega$.
They correspond to the positive and negative helicity states
of the Dirac particle and its anti-particle, which are associated with the
unbroken $Q_1$. We label the two solutions for each $s$ as $\omega^{(\pm)}_{s}$.
With this notation the solutions are
\beq
\label{drmodeli}
\omega^{(\pm)}_{s}(\vec\kappa) = \pm
\left\{\left[\kappa - \frac{s}{2}(\mu_R - \mu_L)\right]^2 +
|m|^2\right\}^{1/2} - \frac{1}{2}(\mu_R + \mu_L)\,.
\eeq
Denoting the particle and anti-particle dispersion relations
by $\omega_s$ and $\bar\omega_s$, respectively, they are to be identified
according to
\beqa
\label{drmodeliantiparticle}
\omega_s(\vec\kappa) & = & \omega^{(+)}_s(\vec\kappa)\nonumber\\
& = & \left\{\left[\kappa - \frac{s}{2}\mu_\phi\right]^2 +
|m|^2\right\}^{1/2} - \frac{1}{2}\mu_{RL}\,,\nonumber\\
\bar\omega_s(\vec\kappa) & = & -\omega^{(-)}_s(-\vec\kappa)\nonumber\\
& = & \left\{\left[\kappa - \frac{s}{2}\mu_\phi\right]^2 +
|m|^2\right\}^{1/2} + \frac{1}{2}\mu_{RL}\,,
\eeqa
where we have used \Eq{murelation}, and defined
\beq
\label{muRLdef}
\mu_{RL} = \mu_R + \mu_L\,.
\eeq
It should be kept in mind that, apart from the explicit dependence
on $\mu_\phi$ in \Eq{drmodeliantiparticle}, $m$ also depends on
$\mu_\phi$ [see \Eq{msymmbreaking}].

\subsection{Discussion}
\label{discussion}

To gain some insight into the solution we can consider some particular
cases. For example, while the particle and anti-particle dispersion relations
are different in general, they are approximately equal in the limit of
small $\mu_{RL}$. We also note that in the limit $\kappa \gg |\mu_\phi|$,
the dispersion relations are approximately independent of $s$.
They are strictly independent of $s$ at $\kappa = 0$,
\beqa
\omega_s(0) & = & \left\{\frac{1}{4}\mu_\phi^2 + |m|^2\right\}^{1/2} -
  \frac{1}{2}\mu_{RL}\,,\nonumber\\
\bar\omega_s(0) & = & \left\{\frac{1}{4}\mu_\phi^2 + |m|^2\right\}^{1/2} +
  \frac{1}{2}\mu_{RL}\,,
\eeqa
which can be interpreted as the effective masses of the particle
and anti-particle.

On top of these effects, the dispersion relations will also get
corrections due to the interactions with the background excitations.
In the context of thermal field theory such corrections can be determined
by calculating the one-loop self-energy diagrams.
As we have already indicated, those calculations are not
in the scope of the present work.

\section{Model II}
\label{sec:modelii}

We consider a massless chiral fermion $f_L$ with an interaction
\beq
L_{\text{int}} = -\frac{\lambda}{2} \phi \bar f^c_R f_L + h.c\,.
\eeq
In this case there is one conserved charge, with
\beq
Q(f_L) = 1\,,\qquad Q(\phi) = -2\,,
\eeq
and the chemical potentials satisfy
\beq
\label{murelatrionii}
\mu_\phi + 2\mu_L = 0\,,
\eeq
where we are denoting by $\mu_L$ the chemical potential of $f_L$.

Proceeding as in \Section{sec:modeli}, the total Lagrangian
is given as in \Eq{modelitotalL}, but in the present case
\beq
\label{L0modelii}
L_0 = \bar f^\prime_L i\lslash{\partial} f^\prime_L + 
\mu_L\bar f^\prime_L\lslash{u} f^\prime_L -
\left(\frac{m}{2}\bar f^{c\,\prime}_R f^\prime_L + h.c.\right)\,,
\eeq
and
\beq
\label{Lintmodelii}
L^\prime_{\text{int}} = -\frac{\lambda}{2\sqrt{2}}
  (\phi_1 + i\phi_2) \bar f^{c\,\prime}_R f^\prime_L + h.c\,.
\eeq
Defining
\beq
f = f^\prime_L + f^{c\,\prime}_R\,,
\eeq
$L_0$ can be written in the form
\beq
L_0 = \frac{1}{2}\bar f\left(i\lslash{\partial} - \Sigma\right)f\,,
\eeq
or in momentum space
\beq
L_0 = \frac{1}{2}\bar f(k)\left(\lslash{k} - \Sigma\right) f(k)\,,
\eeq
where
\beq
\label{Sigmaii}
\Sigma = mL + m^\ast R - \mu_L\lslash{u}L + \mu_L\lslash{u}R\,.
\eeq
Thus in this case, as a consequence of the symmetry breaking, the
fields $f_L$ and $f^c_R$ form a Majorana fermion, with
the two helicities having different dispersion relations.

In order to obtain the solution for the dispersion relation explicitly,
by comparing \Eqs{Sigmaii}{Sigmai} we observe that the equations for
the dispersion relations in the present case can be obtained from those
of Model-I by setting $\mu_R \rightarrow -\mu_L$.
Thus, from \Eq{drmodeli}, making the indicated substitution
and remembering \Eq{murelatrionii} [$\mu_L = -\frac{1}{2}\mu_\phi$],
the solutions in the present case are
\beq
\label{drmodelii}
\omega^{(\pm)}_{s} = \pm
\left\{\left(\kappa - \frac{s}{2}\mu_\phi\right)^2 + |m|^2\right\}^{1/2}\,.
\eeq
Furthermore, by the same identification given in \Eq{drmodeliantiparticle},
in this case we have
\beq
\bar\omega_s(\vec\kappa) = \omega_s(\vec\kappa)\,,
\eeq
that is, the particle and anti-particle dispersion relations are the same,
as it must be for Majorana modes.

Similar to the discussion in \Section{sec:modeli} we can consider
some limiting cases. For illustrative purposes, in
the limit of small or large $\kappa$, the dispersion
relation reduce to
\beqa
\label{drmodeliiexamples}
\omega_s & = & 
  \sqrt{\frac{1}{4}\mu^2_\phi + |m|^2} -
  \frac{\frac{s}{2}\kappa\mu_\phi}{\sqrt{\frac{1}{4}\mu^2_\phi + |m|^2}}
    \qquad \mbox{(small $\kappa$)}\,,
  \nonumber\\[12pt]
  \omega_s & = & \kappa - \frac{s}{2}\mu_\phi +
  \frac{\frac{1}{4}\mu^2_\phi + |m|^2}{2\kappa}
  \qquad \mbox{(large $\kappa$)}\,,
\eeqa
respectively.

\section{Model III}
\label{sec:modeliii}

\subsection{Formulation}

We consider a massive Dirac fermion $f$ with mass $M$, and
an interaction
\beq
L_{\text{int}} = -\frac{\lambda}{2} \phi \bar f^c f + h.c\,.
\eeq
Similar to Model II, there is one conserved charge,
and the chemical potentials satisfy
\beq
\label{mufmuphirelmodeliii}
\mu_\phi + 2\mu_f = 0\,.
\eeq
Putting once more
\beqa
\label{primedfieldsii}
\phi & = & e^{-i\mu_\phi t}\phi^\prime\,,\nonumber\\
f & = & e^{-i\mu_f t} f^\prime\,,\\
\eeqa
instead of \Eqs{L0modelii}{Lintmodelii} in this case we have
\beqa
\label{L0modeliii}
L_0 & = & \bar f^\prime i\lslash{\partial} f^\prime + 
\mu_f\bar f^\prime\lslash{u}f^\prime - M\bar f^\prime f^\prime -
\left(\frac{m}{2}\bar f^{\prime\,c} f^\prime + h.c.\right)\,,\\
\label{Lintmodeliii}
L^\prime_{\text{int}} & = & -\frac{\lambda}{2\sqrt{2}}
  (\phi_1 + i\phi_2)\bar f^{\prime\,c} f^\prime + h.c\,,
\eeqa
where $m$ is given in \Eq{msymmbreaking}.
The mass term $\frac{m}{2}\bar f^{c\,\prime}f^\prime$ breaks the
degeneracy between the two Majorana components of what would otherwise
be a Dirac fermion. $L_0$ in \Eq{L0modeliii} resembles
the kinetic part of the Lagrangian of the pseudo-Dirac
neutrino model\cite{wolfensteinpetcov},
but here it has the additional term involving the chemical potential.

We take $m$ to be complex in general, and denote its phase by
$\theta$, i.e.,
\beq
m = |m|e^{i\theta}\,.
\eeq
To proceed we introduce the Majorana fields
\beqa
f_1 & = & \frac{1}{\sqrt{2}}\left(e^{i\theta/2}f^\prime +
e^{-i\theta/2}f^{\prime\,c}\right)\,,
\nonumber\\
f_2 & = & \frac{1}{i\sqrt{2}}\left(e^{i\theta/2}f^\prime -
e^{-i\theta/2}f^{\prime\,c}\right)\,,
\eeqa
and therefore
\beq
\label{nudecomposition}
f^\prime = \frac{e^{-i\theta/2}}{\sqrt{2}}(f_1 + if_2)\,.
\eeq
In terms of the Majorana fields $f_{1,2}$, $L_0$ becomes
\beq
L_0 = \frac{1}{2}(\bar f_1 i\lslash{\partial} f_1 +
\bar f_2 i\lslash{\partial} f_2) +
\frac{i\mu_f}{2}(\bar f_1\lslash{u} f_2 -
\bar f_2\lslash{u}f_1) - \frac{M}{2}(\bar f_1 f_1 + \bar f_2 f_2)
- \frac{|m|}{2}(\bar f_1 f_1 - \bar f_2 f_2)\,.
\eeq
Therefore, in the absence of the $\mu_f$ term,
$f_1$ and $f_2$ are uncoupled in $L_0$ with masses $M \pm |m|$, respectively.
In the presence of the $\mu_f$ term, $f_1$ and $f_2$ are mixed.
Our purpose now is to obtain the proper combinations that have a
definite dispersion relation in the presence of the $\mu_f$ term.

\subsection{Dispersion relations}

To restate the problem in a more compact algebraic form we introduce
the notation
\beq
f_M = \left(\begin{array}{c}f_1 \\ f_2\end{array}\right)\,.
\eeq
In momentum space, $L_0$ is then
\beq
L_0 = \frac{1}{2}\bar f_M\left(\lslash{k} + \hat\mu_f\lslash{u} -
\hat M\right)f_M\,,
\eeq
where
\beq
\hat\mu_f =
\mu_f\left(\begin{array}{cc}
  0 & i\\
  -i & 0
\end{array}\right)\,,
\eeq
and
\beq
\label{matrixM}
\hat M = \left(\begin{array}{cc}
  M_{+} & 0\\
  0 & M_{-}
\end{array}\right)\,,
\eeq
where we have defined
\beq
M_{\pm} = M \pm |m|\,.
\eeq
The equation for the dispersion relations and the corresponding
eigenspinors is
\beq
\label{eveqmodeliii}
(\lslash{k} + \hat\mu_f\lslash{u} - \hat M)f_M = 0\,.
\eeq

As in the previous cases, we use the Weyl representation of the
gamma matrices, and decompose
\beq
f_i = \left(\begin{array}{c}x_i\chi_s\\ y_i\chi_s\end{array}\right)
  \qquad (i = 1,2)\,,
\eeq
using the helicity spinors $\chi_s$ (defined in \Eq{helicityspinor}) as basis.
The equations for the coefficients $x_i$ and $y_i$ then become
\beqa
(\omega + s\kappa + \hat\mu_f)x - \hat M y & = & 0\,,\nonumber\\
(\omega - s\kappa + \hat\mu_f)y - \hat M x & = & 0\,,
\eeqa
where $x,y$ are two-dimensional spinors in the $f_{1,2}$ \emph{flavor} space,
\beqa
x & = & \left(\begin{array}{c}x_1 \\ x_2\end{array}\right)\,,\nonumber\\
y & = & \left(\begin{array}{c}y_1 \\ y_2\end{array}\right)\,.
\eeqa
Again, if the $\mu_f$ term is dropped, we get back two uncoupled pairs of
equations, in the Weyl representation and the helicity basis,
for two massive fermions with dispersion relations
$\omega = \sqrt{\kappa^2 + (M \pm |m|)^2}$.
We now seek the solutions in the presence of $\mu_f$ term.

Using the first to write
\beq
y = \frac{1}{\hat M}(\omega + s\kappa + \hat\mu_f)x\,,
\eeq
and substituting in the second one, we get the equation for $x$,
\beq
\label{eqx}
\left[(\omega - s\kappa + \hat\mu_f)\frac{1}{\hat M}
  (\omega + s\kappa + \hat\mu_f) - \Hat M\right]x = 0\,.
\eeq
By straightforward algebra, we obtain
\beq
\label{eqxrelation}
(\omega - s\kappa + \hat\mu_f)\frac{1}{\hat M}
(\omega + s\kappa + \hat\mu_f) = \frac{1}{\hat M}\hat A\,,
\eeq
where
\beq
\label{matrixA}
\hat A =
\left(
\begin{array}{cc}
  \omega^2 - \kappa^2 + r\mu^2_f &
  i\mu_f(\omega - s\kappa) + i\mu_f r(\omega + s\kappa)\\[6pt]
  - i\mu_f(\omega - s\kappa) - i\frac{\mu_f}{r}(\omega + s\kappa) &
  \omega^2 - \kappa^2 + \frac{\mu^2_f}{r}
\end{array}
\right)
\eeq
with
\beq
r = \frac{M_{+}}{M_{-}}\,.
\eeq
Substituting \Eq{eqxrelation} in \Eq{eqx} and multiplying by $\hat M$,
the equation for $x$ is
\beq
(\hat A - \hat M^2)x = 0\,,
\eeq
where $\hat M$ and $\hat A$ are given in \Eqs{matrixM}{matrixA},
respectively.

The dispersion relations are obtained by solving the equation
\beq
\label{eqdisprel1}
(A_{11} - M^2_{+})(A_{22} - M^2_{-}) - A_{12}A_{21} = 0\,,
\eeq
where $A_{ij}$ are the elements of the matrix $\hat A$ defined
in \Eq{matrixA}. It follows by inspection of \Eq{matrixA} that the
products of the $A_{ij}$ that appear in \Eq{eqdisprel1} have the form
\beqa
(A_{11} - M^2{_+})(A_{22} - M^2_{-}) & = & \omega^4 + A_1\omega^2 + A_0\,,
\nonumber\\
A_{12}A_{21} & = & A^\prime_1 \omega^2 + A^\prime_0\,,
\eeqa
where $A_{0,1}$ and $A^\prime_{0,1}$ are independent of $\omega$.
\Eq{eqdisprel1} then leads to the following equation for the dispersion
relation,
\beq
\label{eqdisprel2}
\omega^4 - 2b\omega^2 + c = 0\,,
\eeq
where
\beqa
\label{bcmodeliiieq}
b & = & -\frac{1}{2}(A_1 - A^\prime_1)\nonumber\,,\\
c & = & A_0 - A^\prime_0\,.
\eeqa

By straightforward algebra, after some simplifications, we find
\beqa
A^\prime_1 & = & \frac{\mu^2_f}{r}(1 + r)^2\,,\nonumber\\
A^\prime_0 & = & -\frac{\mu^2_f}{r}(1 - r)^2\kappa^2\,,\nonumber\\
A_0 & = & (\kappa^2 + M^2_{+} - r\mu^2_f)
\left(\kappa^2 + M^2_{-} - \frac{\mu^2_f}{r}\right)\,,\nonumber\\
A_1
% & = & -\left[\left(\kappa^2 + M^2_{+} - r\mu^2_f
%  \vphantom{\frac{1}{r}}\right) +
%  \left(\kappa^2 + M^2_{-} - \frac{\mu^2_f}{r}\right)\right]\nonumber\\
%& = & -\left[2(\kappa^2 + M^2 + |m|^2) - \frac{\mu^2_f}{r}(1 + r^2)\right]
%  \nonumber\\
& = & -\left[2(\kappa^2 + M^2 + |m|^2 + \mu^2_f) -
    \frac{\mu^2_f}{r}(1 + r)^2\right]\,.
\eeqa
Then from \Eq{bcmodeliiieq},
\beqa
\label{bcmodeliii}
b & = & \kappa^2 + M^2 + |m|^2 + \frac{1}{4}\mu^2_\phi\,,\nonumber\\
c & = & \kappa^4 + 2\kappa^2\left(M^2 + |m|^2 - \frac{1}{4}\mu^2_\phi\right) +
\left(M^2_{+} - \frac{r\mu^2_\phi}{4}\right)
\left(M^2_{-} - \frac{\mu^2_\phi}{4r}\right)\,.
\eeqa
The dispersion relations are given by
\beq
\label{disprelmodeliii}
\omega^2_{\pm} = b \pm \sqrt{d}\,,
\eeq
with
\beq
\label{dmodeliiieq}
d = b^2 - c\,,
\eeq
where, from \Eq{bcmodeliii},
\beq
\label{dmodeliii}
d = 4M^2|m|^2 + \mu^2_\phi\kappa^2 +
\frac{\mu^2_\phi}{4r}\left[(1 + r)^2(M^2 + |m|^2) +
  2(1 - r^2)M|m|\right]\,.
\eeq
Once again we recall that $m$ is given in \Eq{msymmbreaking}.

\subsection{Discussion}

To gain some insight we can consider various limiting cases.

\begin{description}

\item[Pseudo-Dirac limit.] If the situation is such that the term
$\mu^2_\phi \kappa^2$ in \Eq{dmodeliii} can be dropped (suffciently small
$\mu_\phi$ and/or $\kappa$), then the dispersion relations are given by
\beq
\omega^2_{\pm} = \kappa^2 + M^{\prime\,2}_{\pm}\,,
\eeq
where
\beq
\label{drmodeliiipseudodirac1}
M^{\prime\,2}_{\pm} = M^2 + |m|^2 + \frac{1}{4}\mu^2_\phi \pm \left\{
4M^2|m|^2 + \frac{\mu^2_\phi}{4r}\left[(1 + r)^2(M^2 + |m|^2) +
2(1 - r^2)M|m|\right]\right\}^{1/2}\,,
\eeq
which are the dispersion relations for two fermions
with effective masses masses $M^\prime_{\pm}$.

Further, in the special case that $\mu_\phi$ is sufficiently small that
the explicit $\mu_\phi$ terms can be dropped in \Eq{drmodeliiipseudodirac1}
(while $|m|$ is kept), the dispersion relations reduce to
\beq
\label{drmodeliiipseudodirac2}
\omega^2_{\pm} = \kappa^2 + M^2_{\pm}\,,
\eeq
which resemble the dispersion relations in vacuum for two fermions
with masses $M_{\pm}$, as already anticipated above.
In the neutrino context \Eq{drmodeliiipseudodirac2} is the familiar
pseudo-Dirac neutrino model\cite{wolfensteinpetcov}.
However it must be kept in mind that in the more general
case in which the term $\mu^2_\phi \kappa^2$ in \Eq{dmodeliii} cannot
be dropped, the $\kappa$ dependence of the dispersion relations
does not have the canonical form of
\Eqs{drmodeliiipseudodirac1}{drmodeliiipseudodirac2}.

\begin{figure}
  \begin{center}
\epsfig{file=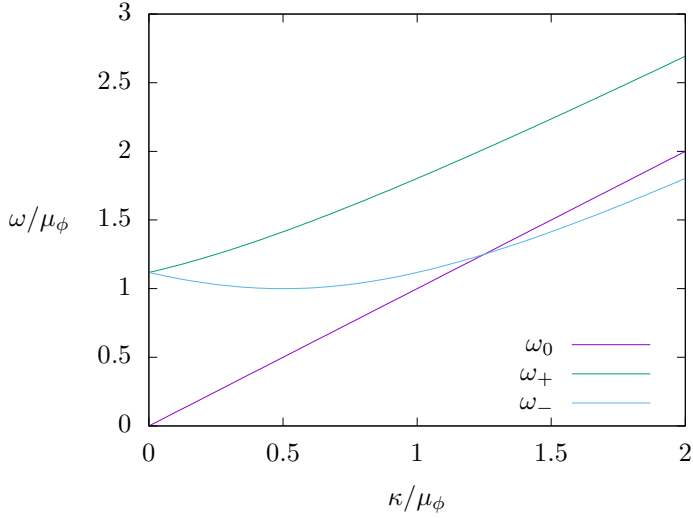,bbllx=163,bblly=274,bburx=474,bbury=517}
\end{center}
  \caption[] {Plot of the dispersion relations of the Majorana modes
    in the case of negligible $M$, given in \Eq{omegamodeliiismallM}.
    For the plot we taken $|m|^2 \sim \mu^2_\phi$.
    For reference, the plot of the dispersion relation $\omega_0 = \kappa$
    is superimposed. 
  \label{fig:wmodeliii}
}
\end{figure}
\item[$|m| \ll M$ limit.] In this limit, the $d$ term in \Eq{dmodeliii}
can be approximated by
\beq
d = 4M^2|m|^2 + \mu^2_\phi\kappa^2 + \mu^2_\phi M^2\,,
\eeq
so that the dispersion relations reduce to
\beq
\label{approxsol1modeliii}
\omega^2_{\pm}(\kappa) = \kappa^2 + M^2 + |m|^2 + \frac{1}{4}\mu^2_\phi \pm
2\sqrt{M^2 |m|^2 + \frac{1}{4}\mu^2_\phi\kappa^2 + \frac{1}{4}\mu^2_\phi M^2}\,.
\eeq
Further, taking the $\kappa\rightarrow 0$ limit,
\beq
\omega^2_{\pm}(0) = \left(M \pm\sqrt{|m|^2 + \frac{1}{4}\mu^2_\phi}\right)^2\,,
\eeq
which can be interpreted as the effective masses of the Majorana modes,
in the $|m| \ll M$ limit. But again, the $\kappa$
dependence of the dispersion relation is different than the one given in
\Eqs{drmodeliiipseudodirac1}{drmodeliiipseudodirac2}.
In the case that $|m|$ can be neglected relative to $\mu_\phi$ (for example,
if $\mu_\phi$ is sufficiently close to $m_\phi$), then \Eq{approxsol1modeliii}
can be approximated by
\beq
\omega_{\pm}(\kappa) = \sqrt{\kappa^2 + M^2} \pm \frac{1}{2}\mu_\phi\,,
\eeq
which resemble the dispersion relation of a neutrino propagating in a matter
background with a Wolfenstein-like potential
$V_{\text{eff}} = \frac{1}{2}\mu_\phi$.

\item[Small $M$ limit.] For sufficently small values of $M$, the
  dispersion relations are approximated by
  \beq
  \label{omegamodeliiismallM}
  \omega^2_{\pm} = \kappa^2 + |m|^2 + \frac{1}{4}\mu^2_\phi \pm
  \mu_\phi\kappa\,.
  \eeq
  Therefore, the two modes have the same efective mass
  \beq
  \omega(0) = \sqrt{|m|^2 + \frac{1}{4}\mu^2_\phi}\,,
  \eeq
  but different dispersion relations away from $\kappa = 0$.
  A plot of \Eq{omegamodeliiismallM} is shown in \Fig{fig:wmodeliii}.

\end{description}

\section{Conclusions and outlook}
\label{sec:conclusions}

In previous works we have carried out a systematic calculation of the neutrino
dispersion relation, as well as the damping and decoherence effects,
when the neutrino propagates in a thermal background of fermions and scalars,
with a Yukawa-type interaction between the neutrino and the background
particles [see \Rref{ns:nuphiresonance} and references therein].

As a complement of that work, the motivation of the present work is to
determine the corresponding quantities for the case in which the
scalar background consists of a Bose-Einstein condensate.
To this end, here we have proposed an efficient and consistent method to treat
the propagation of generic fermions in the background of BE condensate. With
an outlook to possible application in other contexts, we have illustrated
and implemented the method in a general way and not tied to any
specific application. In the present work we have focused exclusively
on the calculation of the dispersion relations.
To model the propagation of the fermions in such an environment,
we assumed some simple Yukawa-type interactions between the fermions
and the scalar. 

As mentioned in the Introduction, the method we use to treat the BE condensate
has been discussed by various authors\cite{weldon:phimu,filippi:phimu,
schmitt:phimu}. In \Section{sec:becmodel} we reviewed
those aspects and details of the method that are relevant for our purposes.
In the following three sections we presented the extension
we propose of that method to treat the propagation of fermions
in the BE condensate, in the context of three generic, but specific,
models of the fermion-scalar interaction.
Specifically in \Section{sec:modeli} we considered two massless
chiral fermions, $f_L$ and $f_R$, with a coupling to the scalar particle
$\phi$ of the form $\bar f_R f_L\phi$ (Model I). In \Section{sec:modelii}
we considered a massless chiral fermion $f_L$ with coupling
$\bar f^c_R f_L\phi$ (Model II). Finally in \Section{sec:modeliii} 
we considered one massive Dirac fermion $f$ with a coupling
$\bar f^c f\phi$ (Model III).

In each case we determined the fermion modes and corresponding dispersion
relations and pointed out some of their particular characteristics. For
example, as a result of the symmetry breaking the propagating mode is
a Dirac fermion and a Majorana fermion in Models I and II, respectively.
In Model III the symmetry breaking produces two non-degenerate Majorana
modes of what otherwise would form a Dirac fermion field in the unbroken phase.
In the latter case, various particular features of the dispersion
relations of the Majorana modes were illustrated by considering
particular limiting cases of the parameters of the model. For example,
one interesting observation is that, while in general 
the two Majorana modes have different effective masses (the value of the
dispersion relation at zero momentum), in some limits the two modes have
the same effective mass although the dispersion relations at non-zero
momentum are different.

The method we propose for the propagation of fermions in a BE condensate
has never being used before, and can be applicable in various contexts,
for example neutrino physics, condensed or nuclear matter systems
and heavy-ion collisions. In addition, the work sets the ground for
considering the case of various fermion flavors, as would be required
for the aplication to neutrinos, or the corrections to the
dispersion relations due to the thermal effects of the background
excitations, that could be required for particular applications.

%\begin{acknowledgments}
The work of S. S. is partially supported by 
DGAPA-UNAM (Mexico) PAPIIT project No. IN103522.
%\end{acknowledgments}

\appendix

\section{Transformation of the chemical potential}
\label{sec:chemicalpotential}

Here we show and state precisely the result we invoke in \Section{sec:becmodel}
regarding the use of the $\phi^\prime$ field (with its corresponding Lagrangian
$L^{(\phi^\prime)}$) in the thermal field theory
calculations while setting $\mu_\phi = 0$ in the partition function.
Moreover, as will become evident, the result also holds
for any other field, not just for the $\phi$ field, that is transformed
in a similar way as we have done for the fermions.

We denote by $Q_A$ the set of conserved charges associated with the
symmetries of the Lagrangian, which are such that
\beq
\label{commQAphi}
[Q_A,\phi] = -q_A\phi\,.
\eeq
The partition function is given by
\begin{equation}
  Z = e^{-\beta H + Q}\,,
\end{equation}
where
\beq
Q = \sum_A \alpha_A Q_A\,,
\eeq
and the chemical potential of $\phi$ is given by
\beq
\mu_\phi = \frac{\alpha_\phi}{\beta}\,,
\eeq
where
\beq
\alpha_\phi \equiv \sum_A \alpha_A q_A\,.
\eeq
From \Eq{commQAphi} we have
\beq
[Q,\phi] = -\alpha_\phi\phi\,.
\eeq

The statement we now show is this: if instead of carrying the calculations with
the $\phi$ field and its original Lagrangian $L$ and corresponding
Hamiltonian $H$, we use the $\phi^\prime$ field,
with the transformed Lagrangian $L^\prime$ and corresponding
Hamiltonian $H^\prime$, then the partition function becomes
\beq
Z = e^{-\beta H^\prime}\,,
\eeq
when it is expressed in terms of the prime field.
As already mentioned, a somewhat exhaustive discussion of this
point is given in \Rref{weldon:phimu}. A simple way to understand this result
is the following.

The evolution equation for $\phi$ is given by
\beq
\label{hamiltoneqphi}
i\partial_t\phi = -[H,\phi]\,,
\eeq
where $H$ is the Hamiltonian corresponding to the Lagrangian $L$.
If now calculate the time derivative of
$\phi^\prime \equiv e^{i\mu_\phi t}\phi$ we get
\beqa
\partial_t\phi^\prime & = &
i\mu_\phi\phi^\prime + e^{i\mu_\phi t}\partial_t\phi
\nonumber\\
& = & i\mu_\phi\phi^\prime + e^{i\mu_\phi t}i[H,\phi]\nonumber\\
& = & i\mu_\phi\phi^\prime + i[H,\phi^\prime]\nonumber\\
& = & -\frac{i}{\beta}[Q,\phi^\prime] + i[H,\phi^\prime]\nonumber\\
& = & i[H - \frac{1}{\beta}Q,\phi^\prime]\,,
\eeqa
or
\beq
i\partial_t\phi^\prime = -[H - \frac{1}{\beta} Q,\phi^\prime]\,.
\eeq
In other words, the Hamiltonian that governs the evolution
of $\phi^\prime$ is $H - \frac{1}{\beta}Q$, or equivalently
the two Hamiltonians are related by
\beq
\label{hamiltoneqphiprime}
H^\prime = H - \frac{1}{\beta}Q\,.
\eeq
Therefore, when we express the partition function in terms
of the field $\phi^\prime$,
\beq
\label{Znomu}
Z = e^{-\beta H + Q} = e^{-\beta H^\prime}\,.
\eeq
That is, in the calculations using $\phi^\prime$
we use the partition function with its Hamiltonian $H^\prime$
and zero chemical potential.

The reason we emphasize here the operator proof of \Eq{hamiltoneqphiprime},
is because in this way it is applicable to any field (e.g., a fermion field),
not involving Lagrangian dynamics arguments, and therefore
the result shown above holds for any field and not a scalar field.
On the other hand, the Lagrangian formulation we carried out in
\Section{sec:becmodel} is the most efficient way to do the dynamics
of the $\phi^\prime$, which makes very straightforward to solve
the evolution equations, rather than starting with the Hamilton equation,
to find the dispersion relations, propagators and wave functions
of the propagating modes.

\section{Scalar modes of the BE condensate}
\label{sec:phi12modes}

In this appendix we complete the discussion of the model
presented in \Section{sec:becmodel} with regard
to the $\phi_{1,2}$ excitation modes of the BE condensate.
To simplify the notation, here we omit the subscript in
the chemical potential, mass and quartic coupling of the $\phi$
and denote them by simply $\mu, m, \lambda$ (without the $\phi$ subscript),
respectively.

\subsection{Lagrangian for the scalar modes}
  
As already mentioned in \Section{sec:becmodel}, the starting point is
to substitute \Eqs{phi}{phi0} in \Eq{Lmu} to obtain the
Lagrangian for the scalar excitations $\phi_{1,2}$. Doing piece by piece,
\beqa
(\partial^\mu\phi^\prime)^\ast(\partial_\mu\phi^\prime) & = & \frac{1}{2}
\left[(\partial^\mu\phi_1)^2 + (\partial^\mu\phi_2)^2\right]\,,\nonumber\\
i[\phi^{\prime\,\ast} (v\cdot\partial\phi^\prime) -
(v\cdot\partial\phi^\prime)^\ast \phi^\prime] & = &
2\mbox{Re}\left( i\phi^\ast v\cdot\partial\phi\right)\nonumber\\
& = & \phi_2 v\cdot\partial(\phi_1 + \phi_0) -
(\phi_1 + \phi_0)v\cdot\partial\phi_2\nonumber\\
& = & \phi_2 v\cdot\partial\phi_1 -
\phi_1 v\cdot\partial\phi_2 - \phi_0 v\cdot\partial\phi_2\,.
\eeqa
The last term is a total derivative and therefore does not contribute
to the action or the equations of motion and can be dropped. Finally,
\beqa
U(\phi^\prime) & = &  -\frac{1}{2}(\mu^2 - m^2)[(\phi_0 + \phi_1)^2 + \phi^2_2]
+ \frac{1}{4}\lambda[(\phi_0 + \phi_1)^2 + \phi^2_2]^2\nonumber\\
& = & U_0 + U_1 + U_2 + U_3 + U_4\,,
\eeqa
where $U_0$ has been defined in \Eq{U0}, and
\beqa
U_1 & = & \frac{\partial U_0}{\partial\phi_0} \phi_1\,,\nonumber\\
U_2 & = & -\frac{1}{2}(\mu^2 - m^2)[\phi^2_1 + \phi^2_2] +
\frac{1}{2}\lambda \phi^2_0\phi^2_2 + \frac{3}{2}\lambda \phi^2_0\phi^2_1\,,
\nonumber\\
U_3 & = & \lambda\phi_0(\phi^3_1 + \phi_1 \phi^2_2)\,,\nonumber\\
U_4 & = & \frac{\lambda}{4}(\phi^4_1 + \phi^4_2 + 2\phi^2_1\phi^2_2)\,.
\eeqa
The terms $U_{3,4}$ give the self-interactions between $\phi_{1,2}$
which we are not interested in at the moment, $U_0$ is an irrelevant
constant, and $U_1 = 0$ when \Eq{phi0} is used. The quadratic part,
using \Eq{phi0} is
\beq
U_2 = \frac{1}{2} m^2_1\phi^2_1 + \frac{1}{2} m^2_2\phi^2_2\,,
\eeq
where
\beqa
m^2_1 & = & - (\mu^2 - m^2) + 3\lambda\phi^2_0\nonumber\\
& = & 2(\mu^2 - m^2)\,,\nonumber\\
m^2_2 & = & - (\mu^2 - m^2) + \lambda\phi^2_0\nonumber\\
& = & 0\,.
\eeqa
In the second line in each equation we have used \Eq{phi0}.
Therefore, the quadratic part of the $\phi_{1,2}$ Lagrangian is
\beq
L^{(2)}_\phi = \frac{1}{2}
\left[(\partial^\mu\phi_1)^2 + (\partial^\mu\phi_2)^2\right] +
\phi_2 v\cdot\partial\phi_1 - \phi_1 v\cdot\partial\phi_2 -
\frac{1}{2} m^2_1\phi^2_1\,.
\eeq
Thus $\phi_1$ and $\phi_2$ are mixed by the $v^\mu$ term.
The next step is to find the propagator matrix of the $\phi_{1,2}$ complex
and determine the modes that have a definite dispersion relation.

\subsection{Dispersion relations for the scalar modes}

Using matrix notation,
\beq
\hat\phi = \left(
\begin{array}{l}
  \phi_1 \\ \phi_2
\end{array}\right)\,,
\eeq  
the Lagrangian, in momentum space, is
\beq
L^{(2)}_\phi(k) = \frac{1}{2}
\hat\phi^\ast(k)\Delta^{-1}_\phi(k)\hat\phi(k)\,,
\eeq
where
\beq
\label{Deltainv}
\Delta^{-1}_\phi(k) = \left(
\begin{array}{ll}
  k^2 - m^2_1 & 2iv\cdot k\\
  -2iv\cdot k & k^2
\end{array}
\right)\,.
\eeq
The classical equations of motion are then
\beq
\Delta^{-1}_\phi(k)\hat\phi = 0\,.
\eeq
The dispersion relations of the eigenmodes are given by the solutions of
\beq
D = 0\,,
\eeq
where $D$ is the determinant of $\Delta^{-1}_\phi$,
\beq
\label{D1}
D \equiv k^2(k^2 - m^2_1) - 4(v\cdot k)^2\,,
\eeq
or,
\beq
\label{D2}
D = (\omega^2 - \kappa^2)(\omega^2 - \kappa^2 - m^2_1) - \alpha^2\omega^2\,,
\eeq
where we have defined
\beq
\alpha = 2\mu\,.
\eeq
The dispersion relations are determined by solving
\beq
\label{dispreleq0}
(\omega^2 - \kappa^2)(\omega^2 - \kappa^2 - m^2_1) - \alpha^2\omega^2 = 0\,,
\eeq
which we write in the form
\beqa
(\omega^2 - \kappa^2)(\omega^2 - \kappa^2 - m^2_1)
- \alpha^2(\omega^2 - \kappa^2) - \alpha^2\kappa^2 & = & 0\,,\nonumber\\
(\omega^2 - \kappa^2)^2 - (\omega^2 - \kappa^2)(m^2_1 + \alpha^2)
- \alpha^2\kappa^2 & = & 0\,.
\eeqa
This is a quadratic equation for $(\omega^2 - \kappa^2)$ with solutions
\beq
\label{omegapm}
\omega^2_{\pm}(\kappa) = \kappa^2 + \frac{1}{2}(m^2_1 + \alpha^2) \pm
\sqrt{\frac{1}{4}(m^2_1 + \alpha^2)^2 + \alpha^2\kappa^2}\,,
\eeq
and obviously,
\beq
\label{D3}
D = (\omega^2 - \omega^2_{+})(\omega^2 - \omega^2_{-})\,.
\eeq
Thus, the masses of the propagating modes are
\beqa
m^2_{+} & = & m^2_1 + \alpha^2\,,\nonumber\\
m^2_{-} & = & 0\,.
\eeqa
The zero mass mode is the realization of the Goldstone mode associated
with the breaking of the global $U(1)$ symmetry.

The corresponding eigenvectors satisfy
\beq
\left.\Delta^{-1}_\phi(k)\right|_{\omega = \omega_{s}}\hat\phi_{s}(\kappa)= 0\,,
\eeq
where $s = \pm$. Writing
\beq
\hat\phi_s(\kappa) = 
\left(
\begin{array}{l}
  a_s \\ b_s
\end{array}\right)\,,
\eeq  
the equations for the components are
\beqa
(\omega^2_s -\kappa^2 - m^2_1)a_s + i\alpha\omega_s b_s & = & 0\,,\nonumber\\
-i\alpha\omega_s a_s + (\omega^2_s - \kappa^2)b_s & = & 0\,.
\eeqa
We write the solutions in the form
\beqa
\label{phisolpm}
\hat\phi_{+}(\kappa) & = & \frac{1}{\sqrt{N_{+}}}
\left(
\begin{array}{l}
  \omega^2_{+} - \kappa^2 \\ i\alpha\omega_{+}
\end{array}\right)\,,\nonumber\\
\hat\phi_{-}(\kappa) & = & \frac{1}{\sqrt{N_{-}}}
\left(
\begin{array}{l}
  -i\alpha\omega_{-} \\ \omega^2_{-} - \kappa^2 - m^2_1
\end{array}\right)\,.
\eeqa
The normalization factors $N_{\pm}$ are determined
by requiring that the one-particle contribution to the propagator from
the eigenmodes coincide with the form of the propagator
near the dispersion relations ($\omega \rightarrow \omega_s$).
The procedure is the following. Instead of expressing $\hat\phi$
in terms of the $1,2$ modes,
\beq
\hat\phi(k) = \left(\begin{array}{c}\phi_1(k) \\ \phi_2(k)\end{array}\right)\,,
\eeq
it is expressed in terms of the modes that have a definite dispersion relation,
\beq
\hat\phi(k) = \sum_{s = \pm} c_s(k)\hat\phi_s(\kappa)\,,
\eeq
where the $\phi_s$ are the eigenvectors found above.
The \emph{free-field} $\hat\phi(x)$ is then expanded in the usual form,
\beq
\hat\phi(x) = \sum_{s = \pm}\int\frac{d^3\kappa}{(2\pi)^3 2\omega_s(\kappa)}
\left[e^{-ik_s\cdot x}a_s(\vec\kappa)\hat\phi_s(\kappa) +
  e^{ik_s\cdot x}a^\ast_s(\vec\kappa)\hat\phi^\ast_s(\kappa)\right]\,,
\eeq
with
\beq
[a_s(\vec\kappa),a^\ast_s(\vec\kappa^\prime)] = 
(2\pi)^3 2\omega_s(\kappa)\delta(\vec\kappa - \vec\kappa^\prime)\,,
\eeq
and
\beq
k^\mu_s = (\omega_s,\vec\kappa)\,.
\eeq

The one-particle contribution to the propagator from a given mode is then
\beq
\label{Deltaphionep}
\left.\Delta_\phi\right|_{\text{one-particle},s} =
\frac{\hat\phi_s \hat\phi^\dagger_s}{2\omega_s(\omega - \omega_s)}\,.
\eeq
For reference and example, we give explicitly the formula for $s = +$,
\beq
\label{phiphidagger}
\hat\phi_{+}\hat\phi^\dagger_{+} = \frac{1}{N_{+}}
\left(\begin{array}{ll}
  (\omega^2_{+} - \kappa^2)(\omega^2_{+} - \kappa^2) &
  (\omega^2_{+} - \kappa^2)(-i\alpha\omega_{+})\\
  i\alpha\omega_{+}(\omega^2_{+} - \kappa^2) &
  (i\alpha\omega_{+})(-i\alpha\omega_{+})
\end{array}\right)\,.
\eeq
On the other hand, by inverting \Eq{Deltainv}, we obtain the propagator
of the $\phi_{1,2}$ complex
\beq
\label{Deltaphi}
\Delta_\phi(k) = \frac{1}{D}
\left(\begin{array}{ll}
  k^2 & -i\alpha\omega\\
  i\alpha\omega & k^2 - m^2_1
\end{array}
\right)\,.
\eeq
where $D$ is given in \Eq{D2}\footnote{%
  To leading order in $\mu$,
  \beq
  \Delta_\phi(k) \simeq
  \Delta^{(1)}_\phi(k) \equiv \left(\begin{array}{ll}
    \frac{1}{k^2 - m^2_1} & \frac{-i\alpha\omega}{d}\\
    \frac{i\alpha\omega}{d} & \frac{1}{k^2}
  \end{array}
  \right)\,,
  \eeq
  where
  \beq
  d = k^2(k^2 - m^2_1)\,.
  \eeq
}.
The propagator has poles at the dispersion relations given in \Eq{omegapm}.
Using \Eq{D3}, near the $\omega = \omega_{+}$ pole, \Eq{Deltaphi} gives
\beq
\label{Deltaphipole}
\left.\Delta_\phi(k)\right|_{\omega\rightarrow\omega_{+}} =
\frac{1}{2\omega_{+}(\omega - \omega_{+})(\omega^2_{+} - \omega^2_{-})}
\left(\begin{array}{ll}
  \omega^2_{+} - \kappa^2 & -i\alpha\omega_{+}\\
  i\alpha\omega_{+} & \omega^2_{+} - \kappa^2 - m^2_1
\end{array}
\right)\,.
\eeq
The normalization factor $N_{+}$ is determined by requiring that
\beq
\left.\Delta_\phi\right|_{\omega\rightarrow\omega_{+}} =
\left.\Delta_\phi\right|_{\text{one-particle},+}\,.
\eeq
Comparing \Eqs{Deltaphionep}{Deltaphipole}, the normalization
factor is then determined by requiring that
\beq
\hat\phi_{+}\hat\phi^\dagger_{+} = 
\frac{1}{\omega^2_{+} - \omega^2_{-}}
\left(\begin{array}{ll}
  \omega^2_{+} - \kappa^2 & -i\alpha\omega_{+}\\
  i\alpha\omega_{+} & \omega^2_{+} - \kappa^2 - m^2_1
\end{array}
\right)\,,
\eeq
and using \Eq{phiphidagger}
(and remembering \Eq{dispreleq0}, for the 22 element)
we then obtain the wave function renormalization factor
\beq
\frac{1}{N_{+}} = \frac{1}{\omega^2_{+} - \omega^2_{-}}
\frac{1}{\omega^2_{+} - \kappa^2}\,.
\eeq
Applying similar arguments to $N_{-}$,
\beq
\frac{1}{N_{-}} = \frac{1}{\omega^2_{+} - \omega^2_{-}}
\frac{1}{\kappa^2 + m^2_1 - \omega^2_{-}}\,.
\eeq

%\bibliographystyle{ieeetr}
%\bibliography{main}

\end{document}